\documentclass{PoS}
\usepackage{amsmath}

\title{Baryon bag simulation of QCD in the strong coupling limit}

\ShortTitle{Baryon bag simulation of QCD in the strong coupling limit}

\author{\speaker{Oliver Orasch} \footnote{This work is partly supported by the Austrian Science Fund FWF, grant I 2886-
N27 and the FWF DK ''Hadrons in Vacuum, Nuclei and Stars''. O.\ Orasch would like to thank Duke University for their 
generous hospitality during a research stay where a large part of this work was conducted. Furthermore, O.\ Orasch thanks the 
Austrian Marshall Plan Foundation for financial support.}\\
        University of Graz, Institute of Physics\footnote{Member of NAWI Graz.}, Universit\"atsplatz 5, 8010 Graz, Austria\\
        E-mail: \email{oliver.orasch@uni-graz.at}}
        
\author{Shailesh Chandrasekharan \footnote{The work of SC is partially supported by the U.S. Department of Energy, Office of Science, 
Nuclear Physics program under award No.~DE-FG02-05ER41368.}\\
        Duke University, Physics Department, Box 90305, 27704 Durham, NC, USA\\
        E-mail: \email{sch@phy.duke.edu}}
        
\author{Christof Gattringer\\
        University of Graz, Institute of Physics, Universit\"atsplatz 5, 8010 Graz, Austria\\
        E-mail: \email{christof.gattringer@uni-graz.at}}
        
\author{Pascal T{\"o}rek\\
        University of Graz, Institute of Physics, Universit\"atsplatz 5, 8010 Graz, Austria\\
        E-mail: \email{pascal.toerek@uni-graz.at}}

\abstract{We explore the possibility of a simulation of strong coupling QCD in terms of so-called \textit{baryon bags}. 
In this form the known representation in terms of monomers, dimers and baryon loops is reorganized such that the baryon 
contributions are collected in space time domains referred to as baryon bags. Within the bags three quarks propagate 
coherently as a baryon that is described by a free fermion, whereas the rest of the lattice is solely filled with interacting 
meson terms, i.e., quark and diquark monomers and dimers. We perform a simulation directly in the baryon bag language 
using a newly developed worm update and show first results in two dimensions.}

\FullConference{The 37th Annual International Symposium on Lattice Field Theory - LATTICE2019\\
		16-22 June, 2019\\
		Hilton Hotel, Wuhan, Hubei, China.}

\begin{document}

\section{Introduction}

In recent years different types of worldline representations of lattice field theories were explored for numerical simulations. 
In this contribution we revisit strong coupling QCD, where the baryonic contributions of the well known monomer, dimer 
and loop representation \cite{Baryons,MDPrep, SCPT}
can be partly resummed into the so-called baryon bag 
representation \cite{BaryonBags}. In this form three quarks propagate coherently like a free quark inside fluctuating
space time domains -- the baryon bags. In the complementary domain outside the baryon bags the degrees of freedom are
quark and diquark monomers and dimers. We discuss this representation and present first results of a numerical simulation
partly using a newly developed worm algorithm.

\section{Baryon bag representation of strong coupling QCD}

We begin with setting our conventions for strong coupling QCD and briefly discuss the baryon bag representation. 
In conventional form the partition function of strong coupling QCD is given by
\begin{equation}
Z  \; = \;  \int \!\! \mathcal{D}\big[\overline{\psi}\psi\big] \int \!\!\mathcal{D}\big[U\big] \;
\text{e}^{\, S_F\left[\overline{\psi},\psi,U\right]} \; .
\label{eq:part_sum}
\end{equation} 
The Grassmann and SU(3) Haar measures are product measures,
$\int \!\! \mathcal{D}\big[\overline{\psi}\psi\big]  =  
\prod_{x} \int \! \prod_{a = 1}^{3}\text{d}\overline{\psi}_{x,a}\text{d}\psi_{x,a}$
and $\int \!\! \mathcal{D}[U] =  
\prod_{x,\nu} \int_{\text{\tiny SU(3)}} \!\! \text{d}U_{x,\nu}$.
We use one flavor of staggered quarks described by the action
\begin{equation}
S_F\left[\overline{\psi},\psi,U\right] \; = \; \sum_x\Big(2m\overline{\psi}_x \psi_x + 
\sum_{\nu} \xi_{x,\nu}\Big[ \text{e}^{\mu\delta_{\nu, d}}\overline{\psi}_{x} U_{x,\nu} \psi_{x+\hat{\nu}} - 
\text{e}^{-\mu\delta_{\nu, d}}\overline{\psi}_{x+\hat{\nu}}U^{\dagger}_{x, \nu} \psi_{x} \Big] \Big) \, ,
\label{eq:stag_act}
\end{equation}
where $m$ is the bare quark mass and $\mu$ the quark chemical potential. The quark fields $\psi_x$ and  
$\overline{\psi}_x$ are three-component Grassmann vectors, with each component representing one of the colors. 
They live on the sites of a $d$-dimensional lattice of volume $V = N_s^{d-1}N_t$, while the SU(3)-valued gauge fields 
$U_{x,\nu}$ live on the links of the lattice. For the fermions, we choose periodic boundary conditions in spatial 
($\nu = 1,\, ... \; d-1$) and anti-periodic boundary conditions in temporal ($\nu = d$) direction while the gauge fields are 
periodic in all directions. Here we are interested in finite temperature which we may vary by changing the temporal 
extent and with a temporal anisotropy $t$ that we introduce for the time direction. 
Together with the staggered sign functions, 
$\gamma_{x, 1} = 1, \; \gamma_{x, 2} = (-1)^{x_1}, \, ... \; \gamma_{x, d} = (-1)^{x_1 + \, ... \, + x_{d-1}}$, we combine 
the anisotropy $t$ in the link factor $\xi_{x,\nu} = t^{\delta_{\nu,d}}\gamma_{x,\nu}$. We remark that 
increasing $t$ corresponds to increasing the temperature \cite{AnisoLat}.

Having fixed the notation, let us now briefly introduce the baryon bag representation. We refrain from giving a full 
derivation of the baryon bag partion sum and refer the interested reader to \cite{BaryonBags} where the mapping 
is presented in detail. In this contribution we provide a different derivation that departs from the worldline representation 
of strong coupling QCD proposed by Karsch and M\"utter and makes clearer the connection between the two representations. The strong coupling QCD partition sum can be exactly rewritten in the form \cite{MDPrep}
\begin{equation}
Z \; = \; \sum_{\{n, d, \ell\}} w_n(m) \; w_d(t) \; w_{\ell}(\mu,t) \; .
\label{eq:Z_KM}
\end{equation}
The degrees of freedom in this representation are monomers $n_x \, \in \, \{0,1,2,3\}$ that are assigned to sites $x$ and 
correspond to mass terms. In addition we may activate dimers $d_{x,\nu} \, \in \, \{0,1,2,3\}$ that correspond to a forward 
hop of a quark followed by a backward hop on the same link $(x,\nu)$. Finally, strong coupling QCD also allows for
baryon loops that are described by fermionic link variables $\ell_{x,\nu} = 0, \pm 1$. The $\ell_{x,\nu}$ must form 
non-intersecting, oriented, and closed fermion loops that correspond to three quarks propagating 
coherently. Admissible configurations of the new variables must satisfy the Grassmann constraints at all sites $x$, i.e.,
\begin{equation}
n_x \, + \, \sum_{\nu} d_{x,\nu} \, + \, \sum_{\nu} \frac{3}{2} |\ell_{x,\nu}| \; = \; 3 \; \; \; \; \forall \; x \;.
\label{eq:GM_const}
\end{equation}
In principle, monomers and dimers carry color. This information, however, can be absorbed in combinatorial factors \cite{Baryons, MDPrep, BaryonBags} that enter the weights $w_n(m)$,  $w_d(t)$ and 
$w_{\ell}(\mu,t)$ in (\ref{eq:Z_KM}). We do not provide the explicit form of these weights and refer the reader to 
 \cite{MDPrep}.

As it stands, the partition sum (\ref{eq:Z_KM}) is not directly suitable for a Monte Carlo simulation. Due to the fermion 
nature of the baryon worldlines, the weights of the baryon loops, $w_{\ell}(\mu,t)$, are not strictly posititive.
Therefore, for a Monte Carlo update we need to group sets of configurations in such a way that the combined 
weights are strictly positive. 

As already remarked, baryon bags can be viewed as a particular resummation strategy of the Karsch-M\"utter 
representation such that real and positive weights emerge.
The first observation for identifying the baryon bags is that one may saturate the 
Grassmann constraint (\ref{eq:GM_const}) using only baryonic 
elements, i.e., 3-monomers ($n_x = 3$), 3-dimers ($d_{x,\nu}=3$) and baryon loops. A baryon bag 
$\mathcal{B}_i$ is then defined as
a domain of the lattice where we sum over all possible configurations of the baryonic elements. Note that the lattice can 
contain several disconnected bags $\mathcal{B}_i$ with $i = 1,2, \, ...$,
and eventually we will sum over all possible configurations of bags. It may be shown 
\cite{BaryonBags} that the baryonic terms used inside the bags can be summed up in a baryon action. 
Consequently, the expansion of the terms in this action generates the baryonic terms inside the baryon bags. 
It turns out that the baryon action $S_B$ is a free staggered action for baryon fields $B_x$ and $\overline{B}_x$,
\begin{equation}
S_B\left[\overline{B},B\right] = \sum_x\Big(2M\overline{B}_x B_x +\sum_{\nu} \xi_{x,\nu}^3\Big[ \text{e}^{3\mu\delta_{\nu,d}}\overline{B}_{x} B_{x+\hat{\nu}} - \text{e}^{-3\mu\delta_{\nu,d}}\overline{B}_{x+\hat{\nu}}B_{x} \Big] \Big) \; .
\label{eq:baryon_act}
\end{equation}
The baryon fields are the composite products of quark fields $B_x = \psi^{1}_x\psi^{2}_x\psi^{3}_x$ and 
$\overline{B}_x = \overline{\psi}^{3}_x\overline{\psi}^{2}_x\overline{\psi}^{1}_x$ with mass $M = 4m^3$. 
It is easy to see \cite{BaryonBags} that the composite baryon fields inherit the 
Grassmann property from the quark fields such that the weight for a bag $\mathcal{B}_i$ is given by a determinant, 
\begin{equation}
\int\limits_{\mathcal{B}_i} \prod_{x\in\mathcal{B}_i} \text{d} B_x \text{d}\overline{B}_x \exp\Big(\sum\limits_{x,y \in \mathcal{B}_i} \overline{B}_x D^{(i)}_{xy} B_y\Big) \; = \; \det D[\mathcal{B}_i]  \; ,
\label{eq:baryon_det}
\end{equation}
where $D^{(i)}_{xy}$ is the free Dirac operator defined by the baryon action (\ref{eq:baryon_act}). Analyzing the 
loop expansion of these bag determinants $\det D[\mathcal{B}_i]$, one may show that they are positive, and for 
non-zero mass this property extends also to small chemical potential.

The union of all bags $\mathcal{B} = \cup_i \mathcal{B}_i$ is denoted as the bag region. 
The rest of the lattice which we refer to as the complementary domain $\overline{\mathcal{B}}$ is filled with mesonic 
contributions that always come with a positive weight. These contributions consist of networks of quark- and diquark 
monomers ($n_x \leq 2$) and quark- and diquark dimers ($d_{x,\nu}  \leq 2$). 
The full partition sum for strong coupling QCD in the baryon bag representation is given by a sum
$\sum_{\{\mathcal{B}\}}$ over all possible baryon bag configurations, 
\begin{equation}
Z \; = \; \sum_{\{\mathcal{B}\}}\prod_{\mathcal{B}_i \in \mathcal{B}} \text{det}D[\mathcal{B}_i] \; \times \; Z_{\overline{\mathcal{B}}} \; .
\end{equation}
$Z_{\overline{\mathcal{B}}} = \sum_{\{n, d || \mathcal{B}\}} w_n(m) \; w_d(t)$ is the weight for the complementary 
domain where $\{n, d || \mathcal{B}\}$ denotes the set of mesonic configurations that are compatible with a given 
baryon bag configuration $\mathcal{B}$.

Two comments are in order here: The baryon bag representation has an interesting connection to the fermion bag 
approach developed by Chandrasekharan \textit{et al.} (see, e.g., \cite{Chandrasekharan:2009wc} for a review). Fermion bags 
are domains of the lattice where the system is described by free fermions while between the bags interaction terms
are used for saturating the Grassmann integral. The picture is similar for the baryon bag representation, with the 
difference that here the free fermions emerge as composite degrees of freedom where three 
quarks propagate coherently as a free baryon. 

Finally, we remark that the baryon bag representation \cite{BaryonBags} allows for new simulation strategies 
compared to the Karsch-M\"utter form \cite{MDPrep}. To attenuate the fermion sign problem, 
Karsch and M\"utter proposed to sample U(3) configurations and then reweight each U(3) configuration to the 
corresponding SU(3) configuration. For the baryon bag representation this detour is not necessary since quantum 
interference of tri-quark monomers, tri-quark dimers and baryon loops yields real and non-negative weights. Thus, the bag 
representation fully takes into account the fermionic degrees of freedom of the model and 
solves the fermion sign problem exactly -- at least for small chemical potential.

\section{Observables and algorithms}

The observables that are conventionally discussed in strong coupling QCD are the chiral condensate and the chiral susceptibility defined as
\begin{equation}
\langle \overline{\psi}\psi \rangle \; = \; \frac{1}{V} \, \frac{\partial \ln Z}{\partial 2m}  \hspace{1cm} \text{and} \hspace{1cm} 
\chi_{\overline{\psi}\psi}  \; = \; \frac{\partial \langle\overline{\psi}\psi\rangle}{\partial 2m} \, + \, 
V\langle \overline{\psi}\psi \rangle^2 \; ,
\end{equation}
where $\chi_{\overline{\psi}\psi}$ is defined to include connected and disconnected terms. In addition, in the baryon bag 
picture another interesting observable is accessible: the average bag size $\sigma_B$ defined as
\begin{equation}
\langle \sigma_B \rangle \; = \; \Big\langle \frac{1}{V} \sum_{\mathcal{B}_i \in \mathcal{B}} |\mathcal{B}_i| \; \Big\rangle,
\end{equation}
where $|\mathcal{B}_i|$ denotes the number of sites in the bag $\mathcal{B}_i$. 
$\langle \sigma_B \rangle$ is a measure for the fraction of the lattice where the physics is described by the 
baryon terms in (\ref{eq:baryon_act}). Loosely speaking, it measures the distribution of fermionic and bosonic 
effective degrees of freedom in the system. In the following, we will always compare a conventional observable -- 
the condensate or the susceptibility -- with the bag size to see how a change in the conventional observable is 
connected to the change of baryonic worldline degrees of freedom. \\

In this work we use two types of algorithms: For a proof-of-concept study in 2D we use a local algorithm that tries to 
exchange a dimer with a pair of monomers and vice versa (see \cite{MDPrep}). Although this algorithm breaks 
down in the chiral limit, it works well with the bag representation. In the following, \textit{bag simulation} refers to this 
strategy.

For cross-checking and simulations on large lattices and higher dimensions we also developed a new worm algorithm. 
Basically, it is a natural extension of the well-known U(3) worm \cite{Adams:2003cca} to arbitrary $m$. We postpone 
a detailed description and proof of detailed balance to future publications \cite{Orasch:2019}, and
only briefly outline the algorithm for general U(N). For starting the worm we pick a site $x$ with a probability of 
$1/V$. Due to the presence of monomers, the worm may start in two ways: With a probability of $n_x/N$ the worm 
starts at the site by naming a monomer \textit{head}. With probability $d_{x,\nu}/N$ the worm decreases $d_{x,\nu}$ 
by $1$. To meet the Grassmann constraint, the algorithm increases $n_x$ by 1 and puts the \textit{head} onto the 
site $x+\hat{\nu}$. For convenience we introduce $t_{\nu} = t^{\delta_{\nu,\pm d}}$ and $w = 2(d-1) + 2t^2 + (2m)^2$. 
Once the worm has started, it exits from the site with a probability $(2m)^2/w$ (by replacing \textit{head} with a monomer) 
or moves with probability $t^2_{\nu}/w$ in the direction $\nu$. By doing so the \textit{head} temporarily moves to an 
intermediate site $x+\hat{\nu}$ by increasing $d_{x,\nu}$ by $1$. The worm has again two options: Exiting this particular 
site with probability $n_{x+\hat{\nu}}/(N-d_{x,\nu})$ by decreasing $n_{x+\hat{\nu}}$ by $1$ or moving the \textit{head} to 
an adjacent site $x+\hat{\nu}+\hat{\mu}$ ($\mu \neq -\nu$) with probability $d_{x+\hat{\nu}, \mu}/(N-d_{x,\nu})$ by decreasing $d_{x+\hat{\nu}, \mu}$ by $1$. 
Thus, the worm shuffles the monomer and dimer configuration by moving around, and by combining the various starting/closing 
steps it is able to raise/lower the monomer number in steps of $2$. Note that despite the fact that we defined a 
\textit{head}, we did not mention a \textit{tail}. In this description any monomer is considered a 
\textit{tail} in the sense that the worm has the option to terminate any time a monomer is encountered.

As described in \cite{Adams:2003cca}, the weights of the worm configurations are defined in such a way that the U(N) 
condensate is simply given by the length $L$ of the worm. To get the corresponding SU(N) observables, we need to 
employ a Karsch-M\"utter-type reweighting strategy. We use \cite{MDPrep}:
\begin{equation}
\langle \mathcal{O} \rangle_{\tiny SU(3)} \; = \; 
\frac{\langle \mathcal{O} \prod_{\ell}(1 + f_{\ell}(t)\text{ sign}(\ell)) 
\rangle_{\tiny U(3)}}{\langle \prod_{\ell}(1 + f_{\ell}(t)\text{ sign}(\ell)) \rangle_{\tiny U(3)}} \; ,
\end{equation}
where $f_{\ell}(t) = 2/(t^{n^{\ell}_t} + t^{-n^{\ell}_t})$ with $n^{\ell}_t = 3N^{\ell}_t - 2N^{\ell}_{Dt}$. $N^{\ell}_t$ is the 
number of loop segments and $N^{\ell}_{Dt}$ is the number of dimers in time direction. In the following, we call this 
strategy \textit{worm simulation}.

\section{Numerical results}

\begin{figure}[t]
\centering
\includegraphics[width=0.7\linewidth]{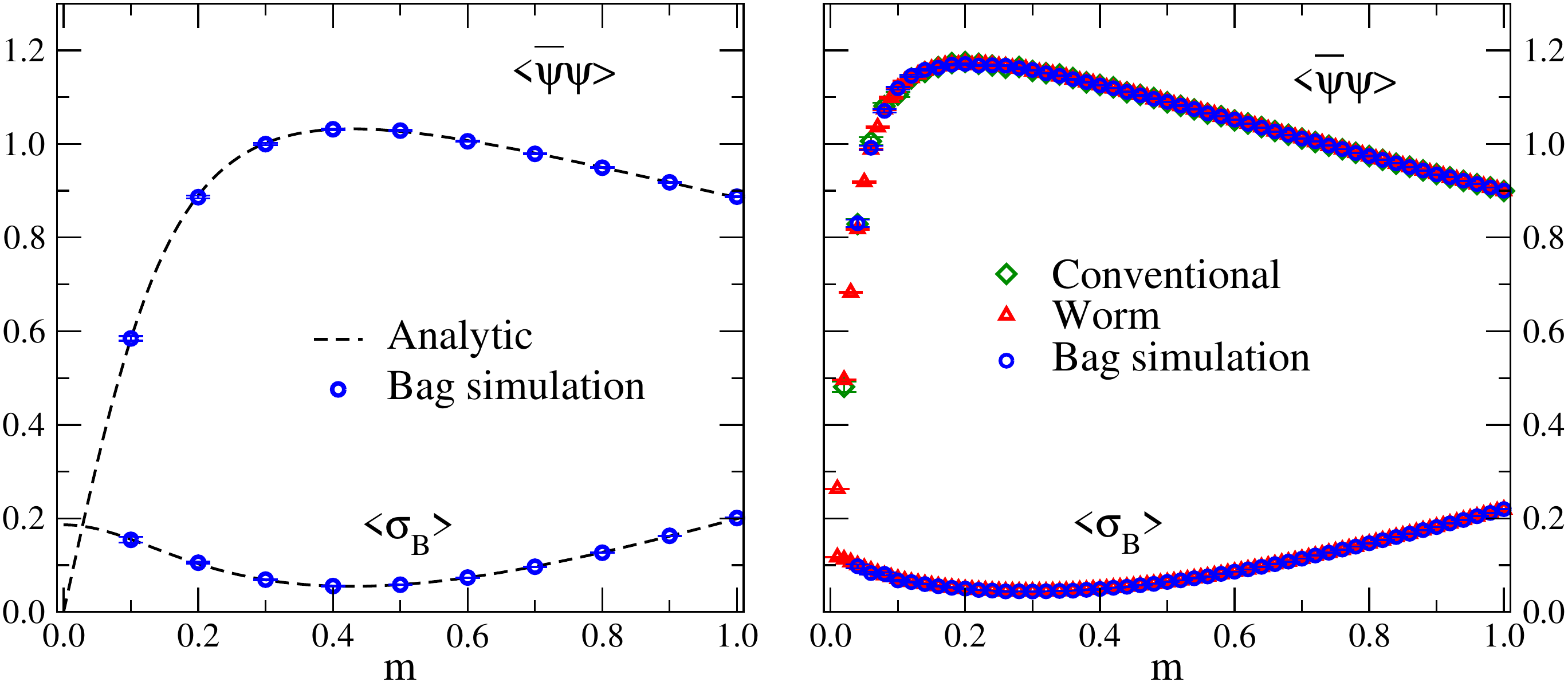}
\caption{Cross-checks of condensate and bag size on small lattices (lhs.: $V = 2\times2$, rhs.: $V = 4\times4$). 
Blue circles represent results from the \textit{bag simulation}. The dashed line is obtained by an exact evaluation. 
Red triangles represent data from the \textit{worm} and green diamonds data from a 
conventional simulation.}
\label{fig:check}
\end{figure}

We begin with presenting the results of a cross-check of the bag representation and different simulation 
strategies on small volumes. The smallest one is $V = 2 \times 2$, where an exact enumeration of the path 
integral is possible (dashed line in the lhs.\ plot of Fig.~\ref{fig:check}). Note, however, that on this smallest lattice 
some configurations that appear on larger lattices are inadmissible such that we also consider a second volume for 
verification ($ V = 4 \times 4$, rhs.\ plot of Fig.~\ref{fig:check}). In Fig.~\ref{fig:check} the results for the bag simulation 
are represented by blue circles while for the worm we use red triangles. For the larger volume, i.e., $V = 4 \times 4$, the 
exact evaluation is not possible, and we use a conventional strong coupling simulation (green diamonds in the rhs.\ plot). 
Fig.~\ref{fig:check} nicely demonstrates that the different representations and simulation strategies give rise to perfectly
matching observables. 

\begin{figure}[t]
\centering
\includegraphics[width=0.6\linewidth]{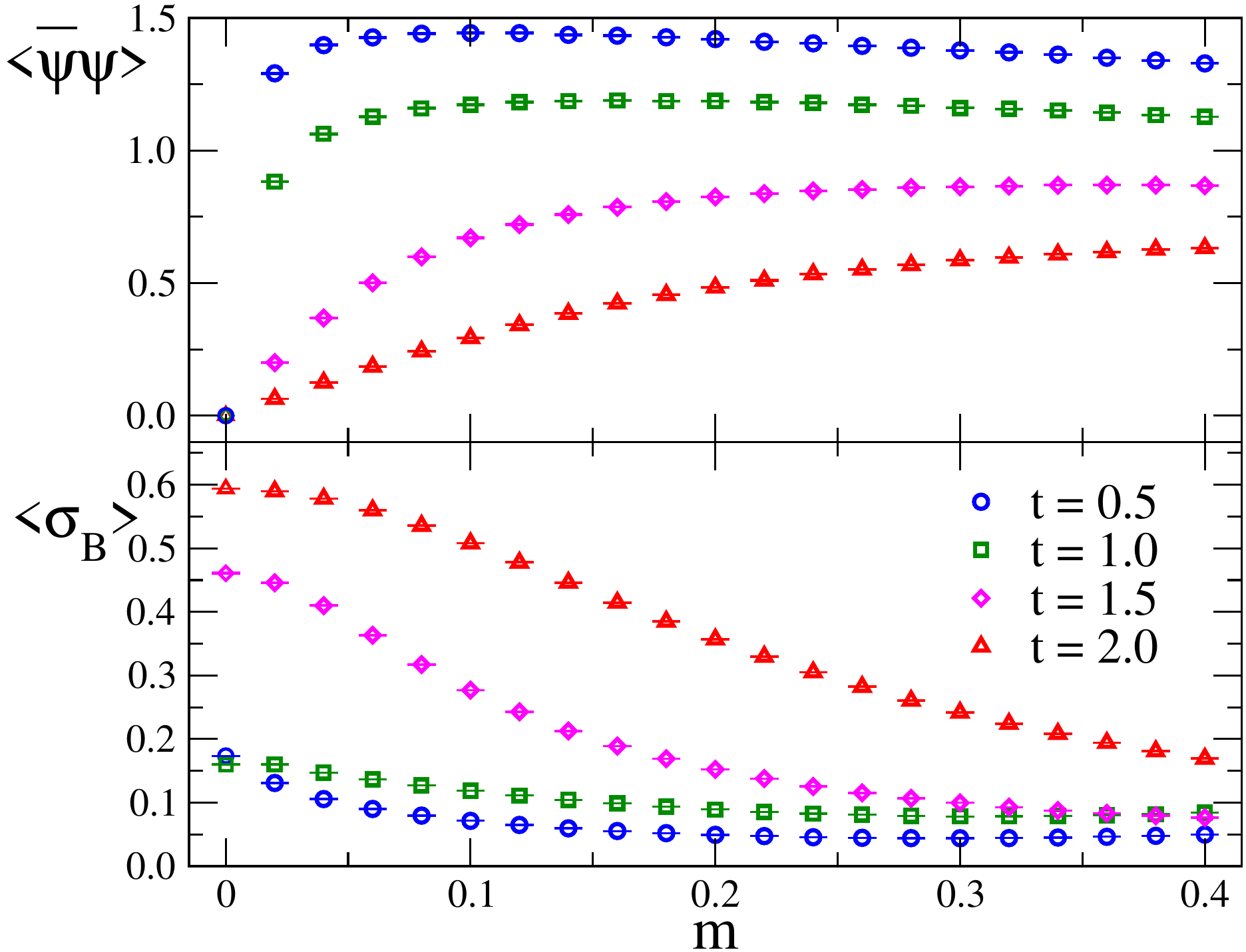}
\caption{Chiral condensate (top) and bag size (bottom) as function of $m$ for different  
anisotropies $t$.}
\label{fig:tfixed}
\end{figure}

We continue with results obtained on $4 \times N_s$ lattices where at $m \neq 0$ we choose $N_s=16$, and for 
$m=0$ we vary $N_s = 8, 12, \, ... \, 24$ to see a potential scaling of the susceptibility. We stress again that here we 
use the \textit{worm} since it is efficient and allows us to study the chiral limit. We use $10^4$ worms for 
equilibration and measure the observables $10^6$ times separated by $10$ worms for decorrelation. 

In Fig.~\ref{fig:tfixed} we study the chiral condensate (top plot) and the bag size (bottom) as a function of the quark mass $m$ 
and compare different values of the anisotropy $t$. For $m=0$ the chiral condensate $\langle \overline{\psi} \psi \rangle$ 
vanishes and then quickly 
increases with $m$ as monomers start to be populated. For larger values of the anisotropy $t$ (i.e., larger temperature) 
this population of monomers competes with temporal link occupation by dimers and baryon loops such that here
the increase of the condensate is slower. The variation of $\langle \overline{\psi} \psi \rangle$ is accompanied by a
variation of the average bag size $\langle \sigma_B \rangle$ where we observe a difference in the $m$-dependence
for the different values of the anisotropy $t$ that in the discussion below we will connect to the phase structure of the system.  

\begin{figure}[t!]
\centering
\includegraphics[height=72mm]{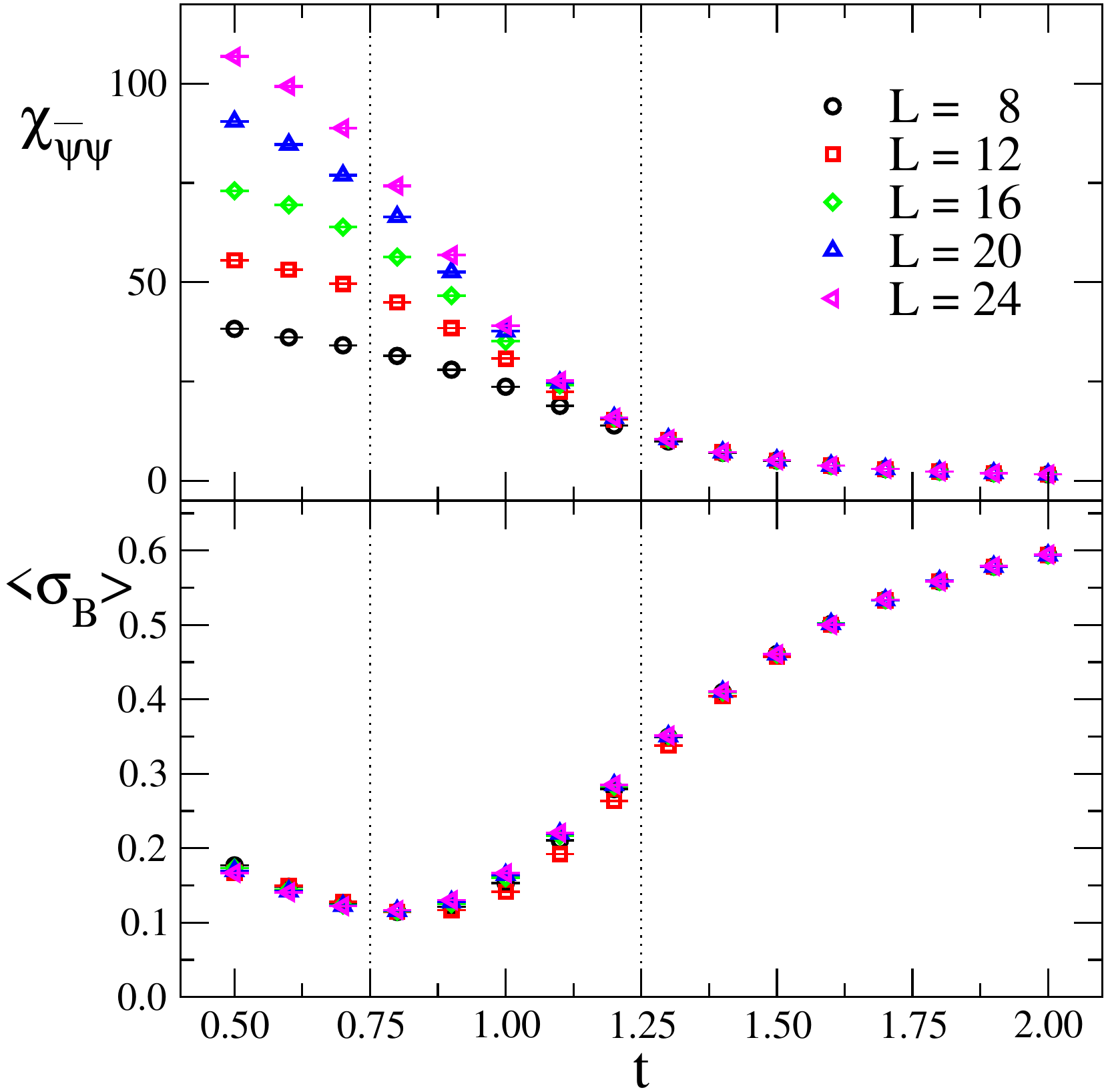}
\hspace{2mm}
\includegraphics[height=72mm]{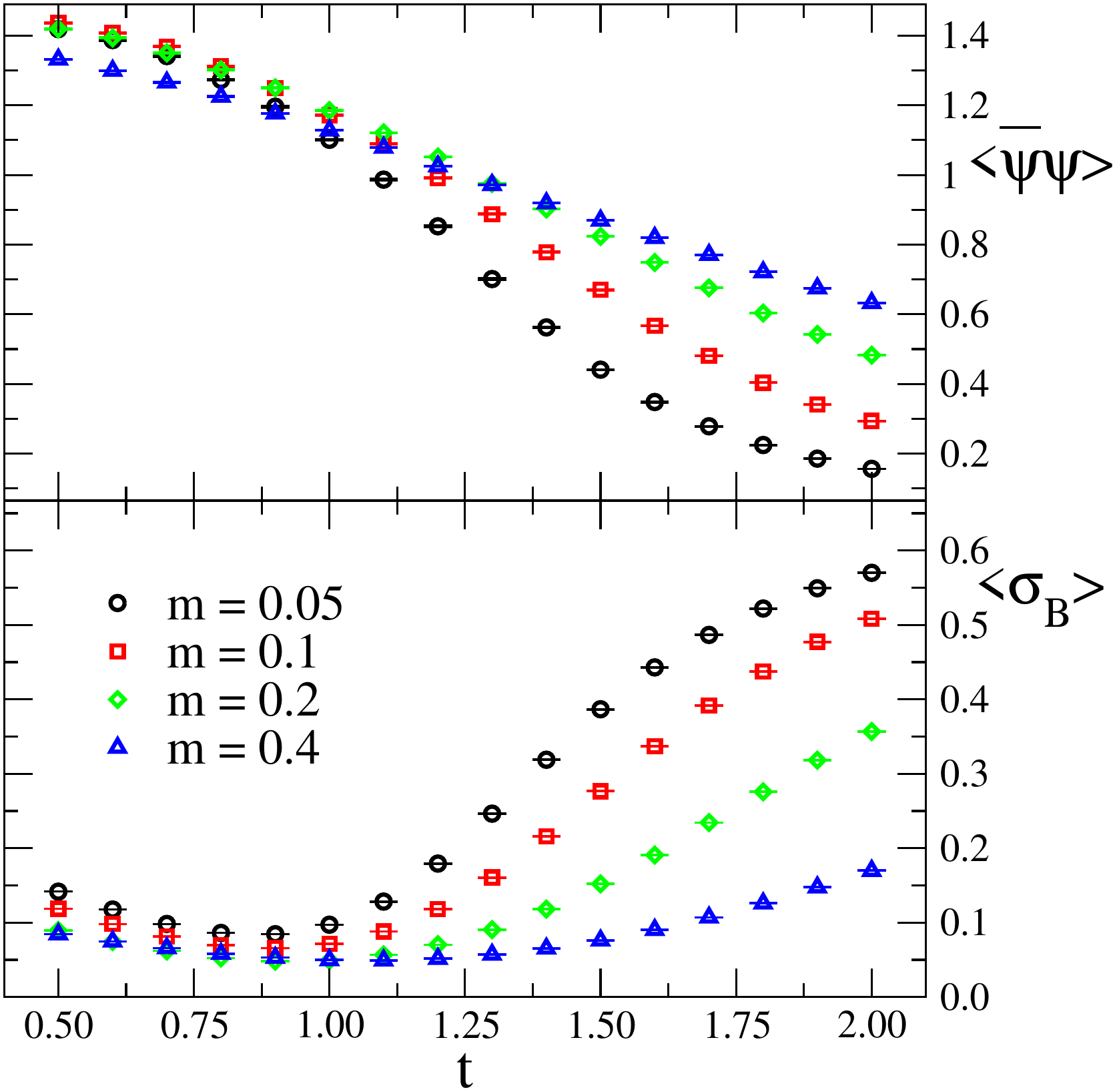}
\caption{Chiral susceptibility (top left) and bag size (bottom left) as a function of $t$ for $m = 0$.
Chiral condensate (top right) and bag size (bottom right) as function of $t$ for different quark masses $m$.}
\label{fig:trun}
\end{figure}

In order to study the $t$-dependence, in Fig.~\ref{fig:trun} we plot the observables now as function of $t$. In the 
lhs.\ plots we study the chiral susceptibility (top) and the bag size (bottom) at $m = 0$ for different volumes while in the 
rhs.\ plots we analyze the chiral condensate (top) and the bag size (bottom) at fixed volume $4 \times 16$ and 
compare different masses. The chiral susceptibility (top left) shows an interesting behavior: for small 
$t$, i.e., low temperature, it shows strong volume dependence while above $t \sim 1.25$ this volume dependence 
is gone -- obviously the system undergoes a change of phase. The corresponding critical behavior has been discussed 
in \cite{Adams:2003cca} and in subsequent work we plan to explore the phase structure in more detail and also address the 
question whether the bag size has a corresponding inflection point near $t \sim 1.25$. When comparing the chiral 
condensate and the bag size as a function of $t$ for non-zero masses (rhs.\ plots) we observe a changing response to
$m$ as a function of $t$. Also this aspect will be addressed in upcoming work. \\

In this contribution we have analyzed strong coupling QCD in the baryon bag representation using a newly 
developed worm algorithm. Our exploratory study of the 2D case so far has focused on testing and verifying the 
new algorithm and a first look at simple observables at different values of the couplings. We are currently working on a 
more detailed analysis of the 2D case with the goal of better understanding the phase structure, in particular in the 
chiral limit. Future work will extend the analysis of strong coupling QCD in the baryon bag representation to 
four dimensions.

\end{document}